# Stable oil-laden foams: Formation and Evolution

Rémy Mensire [*] and Elise Lorenceau[+$]

[*]Univ. Paris-Est, Laboratoire Navier (ENPC-IFSTTAR-CNRS), 2 Allée Kepler, 77420 Champs sur Marne, France

[+]Univ. Grenoble Alpes, LIPhy, Grenoble, France

CNRS, LIPhy, Grenoble, France

[$]elise.lorenceau@univ-grenoble-alpes.fr

Abstract

The interaction between oil and foam has been the subject of various studies. Indeed, oil can be an efficient defoaming agent, which can be highly valuable in various industrial applications where undesired foaming may occur, as seen in jet-dyeing processes or waste water treatment plant. However, oil and foam can also constructively interact as observed in detergency, fire-fighting, food and petroleum industries, where oil can be in the foam structure or put into contact with the foam without observing a catastrophic break-up of the foam. Under specific physico-chemistry conditions, the oil phase can even be trapped inside the aqueous network of the foam, thus providing interesting complex materials made of three different fluid phases that we name oil-laden foam (OLF). In this review, we focus on such systems, with a special emphasis on dry OLF, i.e. with a total liquid volume fraction, $\varepsilon$ smaller than 5%.

We first try to clarify the physical and chemical conditions for these systems to appear, we review the different techniques of the literature to obtain them. Then we discuss their structure and identify two different OLF morphologies, named foamed emulsion, in which small oil globules are comprised within the network of the aqueous foam and biliquid foams, where the oil also comprised in the aqueous foam network is continuous at the scale of several bubbles. Last, we review the state of the art of their evolution in particular concerning topological changes, coalescence, coarsening and drainage.

1. Introduction

Aqueous foams are extraordinary binary systems composed of extremely low cost and simple materials which are air (that represents typically 90% of the foam volume) and foaming aqueous solution. The latter is formed out of water and a tiny amount of interfacial active agent also known as surfactant (usually less than 5% in weight). Aqueous foams are divided up into different elements i) the soap films at the intersection of two bubbles ii) the liquid channels called Plateau borders (PB) at the junction of three soap films and iii) the nodes at the junction of four PB (see Fig. 1a) [1, 2, 3]. The different PB are connected to each other, which allows the liquid phase to flow within the material, hence aqueous foams are often compared to porous media [1, 2, 3]. Nowadays, aqueous foams are inevitable materials which exhibit a wide variety of interesting features such as low density, high specific surface, and original rheological properties. These properties, which have been extensively studied at different lengthscales [1, 2, 3] make foams useful in a wide variety of applications such as in cosmetics or cleaning industries.

In many situations, foams are mixed or put into contact with a third oily liquid phase not miscible with the foaming solution. For example, the low density of foam is used to fight against fire induced by organic materials less dense than water such as fuel. Food processing industry also takes advantage of the high specific surface area of foams which seems to promote the flavors and enhance the food taste to encapsulate various ingredients among which edible oils [4]. Last, foams are also used in soil remediation and decontamination of radioactive tanks to extract more pollutants and oil [86,6]. Indeed, most conventional remediation techniques have drawbacks: cost, area covered by the process, soil excavation, post-treatment of the pollution. However, foams could solve those issues. Foams are a cheap product, easy to transport and generate. If they are able to ingest the organic phase, foam could carry this extracted phase and span more polluted area and invade all the defects and pores thanks to their non-newtonian flow properties and their deformability. Close to soil remediation, the extraction of oil is a research topic of considerable interest for the energy industry. The oil recovery rate remains low around 30 to 40 % with conventional injection techniques. To extract more oil, one of the key ideas is to reduce the mobility of oil. To do so, surfactants can be added to decrease the oil-water interfacial tension and untrap oil from the porous matrix. By injecting carbon dioxide as a volatile organic solvent with a surfactant solution, one create foams that can absorb and carry the oil phase [5].

In these examples, the presence of oil does not dramatically disturb the fragile organization of the air/foaming solution interfaces of the foam which sustain the oil intrusion without catastrophic break-up: the interaction between the foam and the oil is not destructive. While a large number of studies are devoted to destructive interaction between foams and oils (oil is usually considered as an efficient anti-foaming agents), the constructive interaction between liquid phase (non-miscible with the foaming solution) and foams has been less considered. This is mainly due to the ambivalent character of liquid organic phase concerning foams. For example, there is a non-trivial dependency of the oil volume fraction on the foamability: at high oil concentrations, the foamability of the solution may decrease while at lower oil concentrations, the presence of oil may enhance the formation of fresh air-water interfaces and thus the foamability [7, 8, 9, 10, 11]. This example demonstrates the intriguing complexity of the interaction between oil and foam [12].

We try to partially unveil the physics of such systems and consider in this paper Oil-Laden Foams (OLF), which are aqueous foam bounded by air/water interfaces, but containing oil globules. We restrict our studies to "stable" OLF, i.e. systems for which the oil globules do not induce catastrophic foam rupture. We first try to clarify the physical and chemical conditions for these systems to appear and we review the different techniques of the literature to obtain them. Then, we discuss their structure and identify two different OLF morphologies, named foamed emulsion, in which small oil globules are comprised within the network of the aqueous foam and biliquid foams, where the oil also comprised in the aqueous foam network is continuous at the scale of several bubbles. Last, we review the state of the art of their evolution upon topological changes, coarsening, coalescence and drainage.

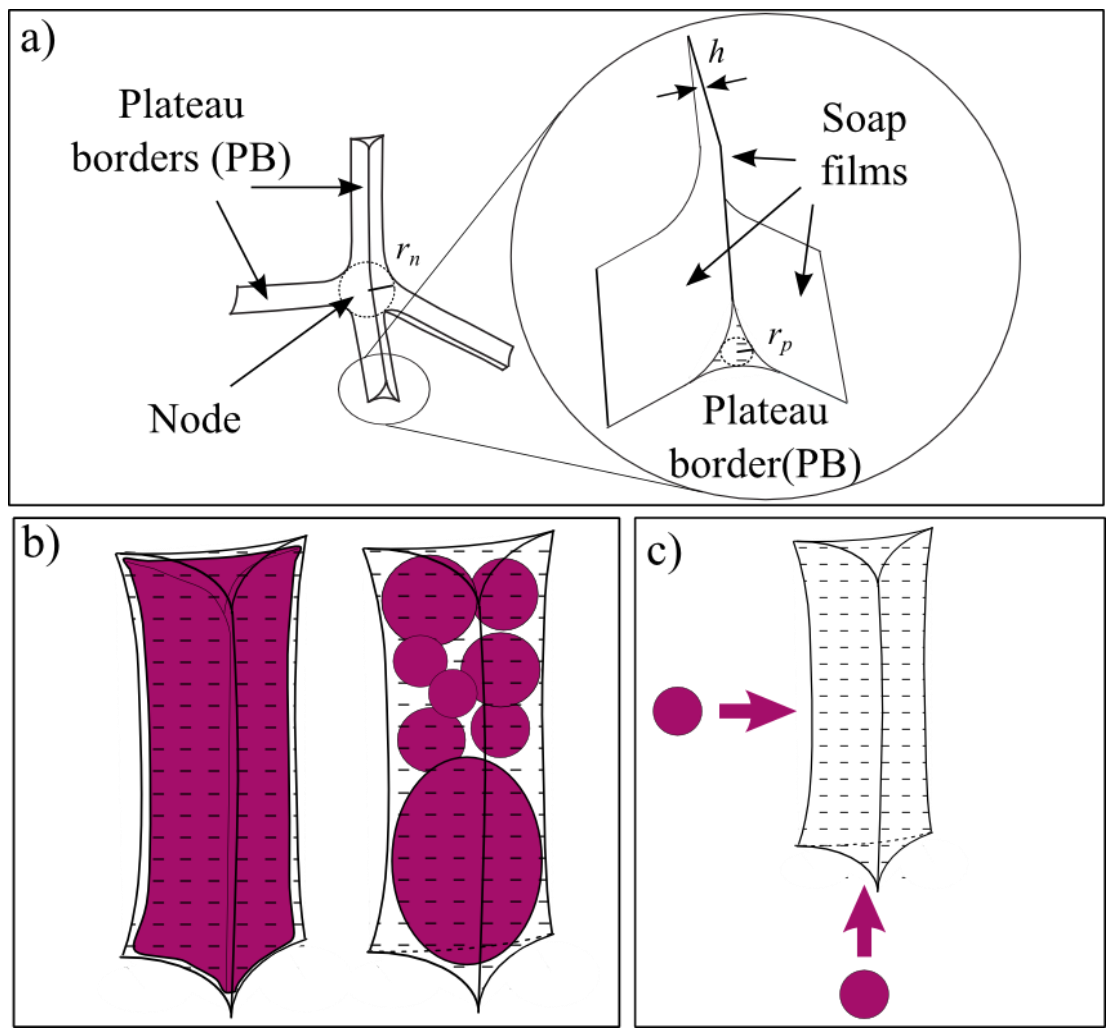

Figure 1 a) the liquid phase of an aqueous foam is divided up into different elements i) the soap films at the intersection of two bubbles ii) the liquid channels called Plateau borders at the junction of three soap films and iii) the nodes at the junction of four Plateau borders. b) Two types of oil-laden foams represented at the scale of a Plateau border. On the left, if the volume of an oil globule is far larger than the PB volume, oil-laden foams where both the oil and the water are continuous phases, can be obtained. We entail this second type of structure biliquid foams. On the right, oil globules of various sizes (in purple) are dispersed in the aqueous network of the foam. Such systems are called foamed emulsion or foamulsion. c) Sketch of the two modes of connection between an oil globule and a PB. The oil globule can enter the PB through the liquid phase or from the gas phase. In that case, it touches the air/liquid interface.

## 2. Production of OLF

The choice of the foaming solution and the oil is critical to obtain "stable" OFL. Indeed, the insertion of oil globules within the network of the aqueous foam should not disturb the fragile organization of the tensio-active molecules at the air/solution interface. Thus, even though it is not the subject of this review, we often refer to the literature devoted to the physics of oil-based antifoaming agents since obtaining stable oil-laden foam can be seen as the complementary goal of destroying aqueous foam with defoaming oil globules. Therefore, in this section, we first review the physico-chemistry conditions required to obtain stable OLF. Then, we detail the experimental set-ups of the literature for the production of OLF.

### 2.1. Oil-Foaming solution compatibility

We first discuss the physico-chemistry conditions required to ensure stable OLF films, then we give the composition of various oil/foaming solution couples of literature which give stable OLF and try to rationalize them.

2.1.1. Physico-chemistry conditions to ensure the stability of OLF films

Here, we discuss the conditions for which an oil globule can be encapsulated within a foam film without breaking it. To do so, we resort to the vast literature devoted to antifoaming processes [8, 9, 10, 13, 14, 15, 12]. We first introduce the classical entry and spreading coefficients (*E* and *S*), which determine whether an oil globule within a foam film is stable or not. They were introduced in the 40's [16, 17] and are presented into great details in [8, 9] introducing $\gamma_{aw}$, $\gamma_{ow}$ and $\gamma_{oa}$, the air-water, the oil-water and the air-oil interfacial tensions. In the following, $\gamma_{aw}$, $\gamma_{ow}$ and $\gamma_{oa}$ refer to the values at equilibrium, i.e. when foaming solution and oil phases have been put into contact for a long time (the interfacial tension values of the uncontaminated foaming solution/air and oil/air can indeed be higher than the values at equilibrium as shown in [18]). As shown in Fig. 2 different situations can appear: the oil globule can stay in the bulk phase or emerge at the air-water interface. This behaviour is controlled by the sign of the entry coefficient *E*, which compares the gain or the loss of interfacial energy between the first and the second configuration. From one configuration to the other, an air-water and an oil-water interface are replaced by an air-oil interface of the same area and subsequently, the entry coefficient is defined as:

$$E = \gamma_{aw} + \gamma_{ow} - \gamma_{ao} \quad (1)$$

An oil globule emerges at the air-water interface if $E > 0$ (loss of interfacial energy). On the opposite, if $E < 0$ (gain of interfacial energy), the globule remains within the film and does not emerge at the air/foaming solution interface. Therefore, this suggests that $E < 0$ is a sufficient thermodynamical condition to obtain a stable oil laden film. However, microscopic observations at the scale of a film showed that the entering and spreading coefficients cannot consistently predict the configuration of the oil at the air-water surface and, consequently, cannot predict the effect of oil on foam stability [19]. The reason for this inconsistency is that this framework does not account for the role of the pseudoemulsion film, which is formed between an air-water interface and the surface of an oil drop which gets closer to it. It is the stability of this film that determines the configuration of the oil drop, as it was earlier suggested in [19, 20].

For thin films, this stability is also linked to the steric or VanderWaals short range forces that build up due to the interaction between the surfactants stabilizing the oil/water and the water/air interfaces. Considering this constraint, Bergeron and coworkers incorporated the thin-film forces into the classical entering and spreading expressions via the knowledge of the disjoining pressure isotherms and established a good correlation between the asymmetric film stability and the foam stability in the presence of oil. [21, 22, 23].

Various geometries have been proposed to quantify these thin-film forces either by using a variant of the porous plate method [21] originally developed in [24, 25] or by measuring the rupture pressure of the asymmetric film, which is formed on the surface of an oil (compound) globule, blown from a capillary tube [15].

In the same framework, it has been shown that the capillary pressure of the air-water interface at the moment of the oil drop entry can be used as a quantitative characteristic of the entry barrier in relationship to the antifoaming efficiency [26, 14, 10, 13, 11]. To measure this pressure, a vertical capillary, partially immersed in a surfactant solution containing oil drops, is held close above the bottom of the experimental vessel. The air pressure inside the capillary is increased, and the water-air meniscus in the capillary is squeezed against the glass substrate until the asymmetric film formed between an oil drop and the solution surface ruptures. Thus, an oil globule in the foam film and exhibiting $E > 0$ enters the foam film surfaces only if the pressure P exerted on the pseudo-disjonction film exceeds a critical value denoted $\Pi$. This suggests that as long as the pressure exerted on the oil globules by the thin film is not too high, oil globules can be encapsulated within a foam film even for a positive value E.

In the same framework, it has been shown that the capillary pressure of the air-water interface at the moment of the oil drop entry can be used as a quantitative characteristic of the entry barrier in relationship to the antifoaming efficiency [26, 14, 10, 13, 11]. To measure this pressure, a vertical capillary, partially immersed in a surfactant solution containing oil drops, is held close above the bottom of the experimental vessel. The air pressure inside the capillary is increased, and the water-air meniscus in the capillary is squeezed against the glass substrate until the asymmetric film formed between an oil drop and the solution surface ruptures. Thus, an oil globule in the foam film and exhibiting E > 0 enters the foam film surfaces only if the pressure P exerted on the pseudo-disjonction film exceeds a critical value denoted Π. This suggests that as long as the pressure exerted on the oil globules by the thin film is not too high, oil globules can be encapsulated within a foam film even for a positive value E.

These two energy barrier approaches quantified in terms of disjoining or critical pressures, which both reflects the existence of short range forces explains why stable OLF can be obtained even for E > 0 as can be seen in Table 1. The different fates of an oil globule are synthesized in Fig. 2: When the oil globule connects the foam film via the air phase, the sign of the entry coefficient determines whether it remains in the bulk of the film or not. Indeed in that case, the long range forces cannot build up because a pseudoemulsion film does not form. When the oil globule enters the foam film via the liquid phase, the oil globule with E < 0 or (E > 0 and P < Π) remains encapsulated in the foam film. In the other cases, the oil globule pierces the air/liquid interface. Having E < 0 is therefore a sufficient condition to obtain stable oil-laden foam films.

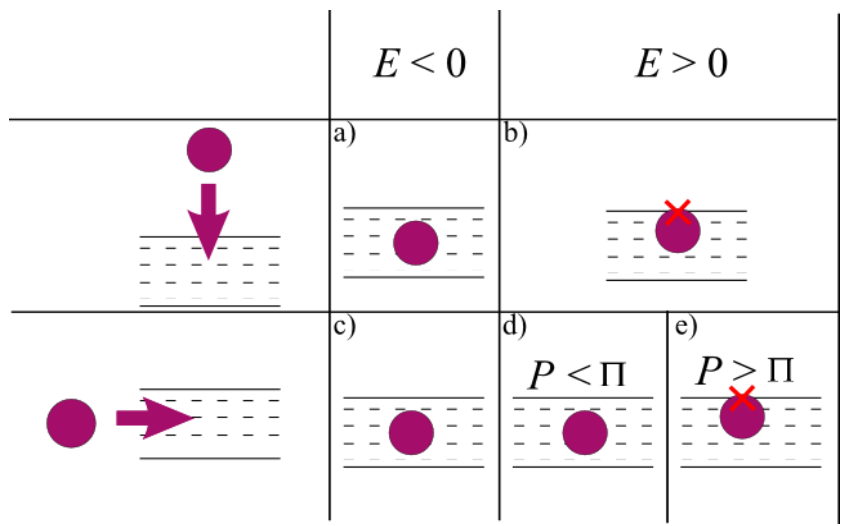

Figure 2 Fate of an oil globule enclosed in the thin film of an aqueous foam. When the oil globule connects the foam film via the air phase, the sign of the entering coefficient determines whether it remains in the bulk of the film a) or not b). In that case, there is no pseudoemulsion film, thus the long range forces which create additional entry barrier do not build up. When the oil globule enters the foam film via the liquid phase, oil globule with E < 0 c) or (E > 0 and P < Π) d) remains encapsulated in the foam film. In the other cases b) and e), the oil globule pierces the air/liquid interface.

We now discuss the fate of the oil globules that have pierced the air/water interface (situations b) and e) depicted in Figure 2). Such oil globules can spread or not. The spreading behaviour depends on the spreading coefficient S which compares the difference of interfacial energy between a naked film and a film with oil at its air-water interface. From the non-spreading to the spreading configuration, the oil globule loses an air- water interface but gains an oil-water and an air-oil interface of the same area. The spreading coefficient S is thus written as:

$$S = \gamma_{aw} - \gamma_{ow} - \gamma_{ao} \qquad (2)$$

If S > 0, it is energetically favorable for the oil globule to spread onto the air- water interface and induces film breaking via Marangoni effects [8, 9]. If S < 0, the oil globule is unlikely to spread. The question that now comes into play is the following: can this stable oil lens induce the break-up of the film? When

E > 0 and S < 0, the oil globule emerges at the air-water interface but does not spread. As liquid foam films usually thin with time due to drainage, this oil globule will eventually emerge at the second air/water interface of a foam film and will bridge it. As proposed by Garrett et al. [8], an oil bridge is stable if the balance of capillary forces at the triple-phase contact line is satisfied, as well as the pressure equilibrium beyond the oil-water and the air-oil interfaces. They discriminate between stable and unstable bridges from the value of a third thermodynamical coefficient B:

$$B = \gamma^2_{aw} + \gamma^2_{ow} - \gamma^2_{ao} \qquad (3)$$

For E > 0 and S < 0, the outcome of the oil-foam film interaction is dictated by the value of B. Foam films bridged with the oil globule are unstable when B > 0, but stable when B < 0.

However comprehensive it might be, we emphasize that to the best of our knowledge, stable oil-laden foams, presenting oil globules piercing the air/water interfaces and bridging the two sides of foam films have never been reported experimentally. Having E < 0 or E > 0 but with P < Π therefore seems to be a mandatory condition to prevent the oil globules to pierce the air/foaming solution interface. We also stress that, in most of the experimental works concerning stable oil-laden foams, the oil globules are not (or only temporarily [27]) observed in the foam films [12], as can be seen in Fig. 3. Due to foam drainage (i.e. liquid flow downwards owing to gravity), the foam films get shrunk and the oil drops get squeezed. This can induce oil emergence in the PBs, where they can accumulate as described below. This is all the more true than the radius of the oil globule is large compared to the film thickness, which is at most in the order of 5 µm. However, when these two lengths are of the same order of magnitude, some oil globules may be trapped within the aqueous foam films [28].

2.1.2. Which oil/foaming solution?

Various compositions of oil/foaming solution couples that give stable OLF are given in the literature. In Table 1, we review some of them and give the main compounds of the foaming solution, the oil types and the value of E when specified in the original work.

Different features can be highlighted from Table 1.

First, as explained in section 2.1.1, a positive values of the entry coefficient, E, can also lead to stable OLF films due to entry barrier effects. Then, organic oil such as rapeseed oil, olive oil or sunflower oil with foaming solution composed of classical anionic surfactants (such as SLES or SDS) often exhibit negative value of E, which make them good candidates for OLF. To obtain a stable OLF, the surfactant should also be largely in excess to compensate for the surfactant dilution by the oil and surfactant adsorption on the oil globules [29, 26]. The critical concentration of surfactant required to obtain stable OLF has been rationalized as a function of oil concentration [29].

For other oils, such as alcanes or silicon oils, E is usually not negative. Yet, stable OLF can be obtained using a mixture of surfactant and co-surfactant such as Myristic Acid or Palmitic Acid, also known as "foam booster"[26, 14]. In particular, the foam stability clearly increases with the relative concentration of foam booster: above a certain relative concentration of betaine, the foams are stable and do not decay within the experimental timescale [26, 14,30]. Why foam booster increases the stability of OLF is not perfectly known. It is generally accepted that the long-range, positive, and repulsive due to electrostatic repulsion of the charges on the interfaces bounding the film stabilizes the pseudoemulsion film [5].

Table 1 Examples of composition of foaming solution/Oil giving stable OLF.

| Foaming Solution | Oil | E (mN/m) |
|---|---|---|
| CAPB+SLES+Mac [31] | Sunflower Oil | -4±2 |

| CAPB+SLES+Mac [31] | Olive Oil | -4±2 |
| CAPB+SLES [32] | Sunflower Oil | -2±2 |
| SDS [29] | Rapeseed Oil | 3±3 |
| NaDBS [12] | Benzene | -0.9 |
| SDS [21] | Tetralin | 3.3 |
| AOS [33] | Tetralin | -1 |
| SDS+NaCl [21] | Tetralin | -4.3 |
| AOS [33] | Acetophenone | -6 |
| NaDBS [12] | Decane | 5.1 |
| AOT+NaCl [34, 18] | C 16 Alcane | -0.2±0.2 |
| AOT+NaCl [34, 18] | C 12 Alcane | -0.1±0.2 |
| AOT+NaCl [34, 18] | C 11 Alcane | 0.3±0.2 |
| ZonylFSK [21] | Dodecane | -0.2 |
| TTAB [28] | Dodecane | Not provided |
| TTAB+Dodecanol [28] | Dodecane | Not provided |
| AOT+NaCl [34, 18] | Squalane | 0.0 |
| CAPB+SLES+Mac [35] | Silicon Oil | 13±2 |
| FS500 [36, 37] | Crude Oil | -3.4±0.2 |
| Fluorad FC75 [38, 39] | Various crude Oil | -3.4±0.2 |
| Fluorad FC75+Brine [38] | Crude Oil | -3.5±0.2 < E <-0.5 ±0.2 |
| Mac+Pac+KOH [30] | Various Oils | Not provided |
| ZonylFSK [33] | Crude Oil | -5 |

## 2.2. Techniques of OLF production

OLF are double dispersed materials constituted of two interconnected biphasic dispersions sharing the same continuous phase: the oil dispersion is constituted of the oil globules in the foaming solution, while the gas dispersion is constituted of the bubble dispersed in the foaming solution. Therefore to produce OLF, one can think of different means. Producing first the oil dispersion (oil globules/foaming solution), then producing the gaseous one (air/foaming solution) by foaming the oil dispersion. This technique, called "foaming an emulsion", typically produces foamed emulsion. The second technique consists in producing first the gas dispersion (bubble/foaming solution) then incorporating the oil inside. This technique, refers to as "Adding oil to foam", can be used to produce biliquid foams. A last technique consists in producing the two dispersions separately and mixing them afterwards.

We don't describe the most commonly used techniques for the production of the initial emulsion or foam, which have been reviewed elsewhere [40]. In the following, we only focus on the different techniques used to entangle these two dispersions together.

### 2.2.1. Foaming an emulsion

To produce a foam out of a continuous phase (here an emulsion), one can nucleate the bubbles from a phase transition or use mechanical means. In all cases, a foamed emulsion as depicted in Fig. 1b) and shown in Figure 3 is obtained.

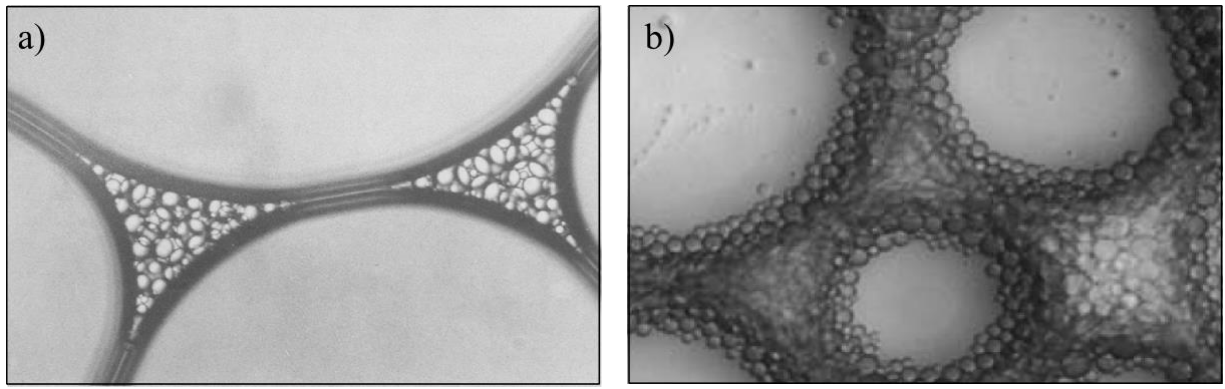

Figure 3 Foamed emulsion (Oil-laden foams with $a < r_p$). Because of the stable pseudoemulsion film, the oil globules drain along with the aqueous phase from the films into the PB. a) The radius of the oil globules is typically 50 μm. Figure reproduced with the permission from [12]. Copyright, API, 1992. b) The radius of the oil globules is typically 10 μm. Figure reproduced with the permission from [29]. Copyright, RSC, 2012.

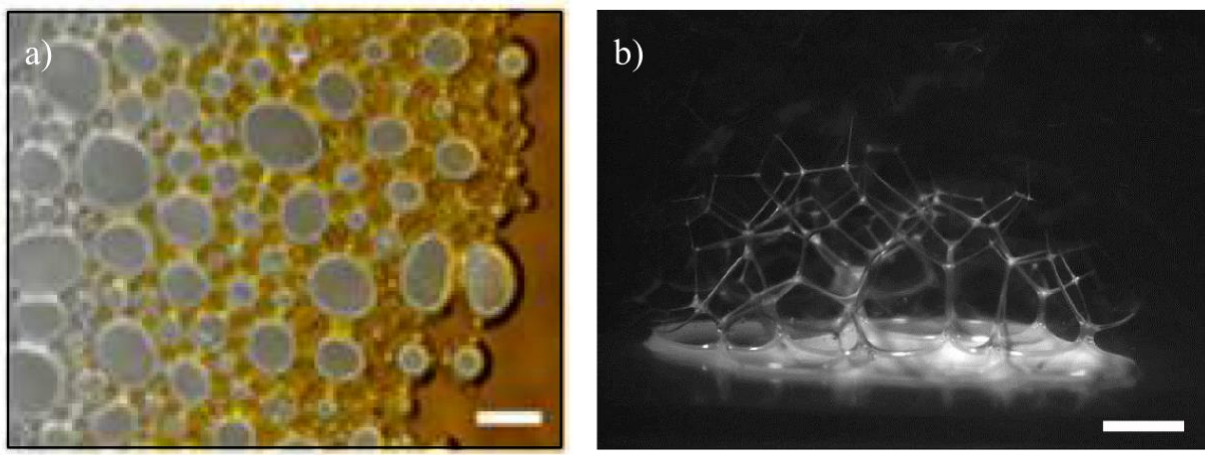

Figure 4 Biliquid foams (Oil-laden foams with $a^3 \gg R\,r_p^2$). a) Imbibition of oil into a 2D foam to 2 s after contact. The white bar represents 200 μm. Figure reprinted with the permission from [30]. Copyright,, ACS, 2014. b) Imbibition of a 3D foam from an horizontal oil slick. For a clearer visualization, a fluorescent dye has been added to the oil phase. When illuminated between a blue filter and a yellow collection filter in front of the camera objective, oil appears bright on the images and water is dark. The white bar represents 2 mm. Figure reprinted from [32].

**Bubble nucleation in emulsion**

The most famous realization of this is the whipped cream charger. In this case, nitrous oxide, which is filled under high pressure in a charger, is released in a cream, which can be constituted of any culinary emulsion [40]. This technique is used to produce lot of aerated food [4, 41].

Bubble nucleation can also be more explosive. This can be seen in mining industries or in natural processe; for example, the sudden explosive degassing of magma in volcanoes can be disastrous. A common example of explosives of mining industry is an emulsion of supersaturated aqueous $NH_4NO_3$ in oil. To make these emulsions detonable, the aqueous drops are loaded with dissolved or small bubbles of Nitrogen. Then, the detonation is initiated by blasting a small amount of another explosive, which generates a shockwave. This adiabatically compresses the nitrogen bubbles allowing them to act as nucleation points, and propagates the detonation. To get a clearer view on this phenomenon, the nucleation of chemically generated nitrogen gas bubbles in oil in water emulsion, has been investigated. The authors of [42] show that OLF can be produced this way, provided that water-soluble surfactants are added to the aqueous phase to inhibit the formation of bubble in the oil phase. Without those, nitrogen bubble nucleation occurs in both the aqueous and oil phases, despite the nitrogen production reaction being a purely aqueous phase process.

**Mechanical foaming of an emulsion**

Various methods have been employed to mechanically produce a foam out of a solution constituted of an emulsion. The major difficulty lies in the fact that the emulsion, if the oil is highly dispersed into fine droplets, can be a non-Newtonian fluid exhibiting various properties such as shear thinning or yield stress. Gas is usually inserted into the emulsion more or less energetically. It can be gently blown out of a needle or a porous glass frit in the emulsion. This technique produces nearly monodisperse bubbles typically ranging between 100 μm and 1 mm with a standard deviation in the bubble size around 5% [28]. The gas fragmentation can also be enhanced by pushing simultaneously the gas and the emulsion in a narrow junction. This microfluidics technique, originally used to obtain single dispersion, has been recently adapted when the continuous phase is a yield stress fluids such as a concentrated emulsion [43, 44]. Higher bubble throughput than gas blowing can be achieved, yet with a less narrow distribution in bubble size (typically with a standard deviation between 5 and 10 %). The foaming can also be induced more energetically using high pressure turbulent mixing method, where the gas and the emulsion are injected at high pressure into a T-junction module. With this method, extremely high throughput can be achieved at the price of a wider distribution in bubble size [29]. Therefore, the choice of one or the other technique must be guided by a compromise between the desired bubble size distribution and the production throughput.

2.2.2. Adding oil to foam

The continuous phase of a pure aqueous foam is in depression compared to the bubble due to the curvature of the gas/liquid interface. This pressure drop, $\Delta P$, can be evaluated using Laplace law which ensures $\Delta P = \gamma_{aw} C$, where $\gamma_{aw}$ is the air/water interfacial tension and C the mean curvature of the interface. For dry foams with a liquid fraction $\varepsilon \sim 1\%$ and millimetric diameter bubbles, $\Delta P=2000$ Pa. Thus, when the pressure in the bubble is at the atmospheric pressure, $P_0$, the pressure in the continuous phase is $P_0 - \Delta P$. This is ensures that when the continuous phase of a dry purely aqueous foam is put into contact with a bath of any liquid, the liquid invades the foam until the equilibrium between capillary and gravity is reached. Oil is absorbed by the aqueous foam thus forming biliquid foams as depicted in Fig. 1a) and shown in Fig. 4. Of course, this is only true if the addition of the liquid does not induce film break-up, in other words, if the physico-chemistry conditions depicted in 2.1.1 are fulfilled. Therefore, two situations can occur:

- If the oil/foaming solution has been chosen with $E < 0$, the oil is instantaneously wetted by the liquid phase thus assuring the stability of the air/foaming solution interface. In that case, when oil - not preemulsified- is added to a dry foam, it invades the foam until a capillary – gravitary equilibrium is reached. This has been depicted in various geometry: invasion of a single PB [38, 39, 35], of a single foam cell [27], a 2D foam [38, 39, 30] or a 3D foam either injecting the oil from a point source [45] or a slick [31].
- If the oil/foaming solution has been chosen with $E > 0$ and a high $\Pi$, then the oil can be successfully added to the foam, only if the connection is done throughout the liquid phase: the oil should not touch the air/foaming solution interface as sketched in Fig. 1c) [27].

2.2.3. Mixing an emulsion and a foam

The foamed emulsion can also be prepared by mixing the emulsion with a separately produced monodisperse foam sharing the same continuous phase than the emulsion. This has been tested in various geometries [46, 47, 48, 49] with usually a relatively high liquid fraction of OLF - typically larger than 10 %. Yet, dedicated emulsion foams with unequaled control of all parameters, i.e. bubble size, gas volume fraction, and concentration of the interstitial emulsion were been obtained very recently [50]. In that case, the mixing has been operated using a mixing device based on flow- focusing method, which allowed to tune the flow rates of both the foam and the emulsion as well as introducing additional foaming solution in order to dilute the emulsion if required.

3. Structure of oil-laden foam

In this section, we discuss how oil globules can be included inside the aqueous network of a foam without destroying its structure. We successively consider the fate of an oil globule inside the different elements of the foam, namely, the soap films, the PB and the nodes. We denote as $v$ and $a$ the volume and the radius of the oil globule defined as $v = 4/3\pi a^3$, $\Omega$ and $R$ the mean volume and the mean radius of the bubbles of the foam defined as $\Omega = 4/3\pi R^3$. We recall that $\varepsilon$ is the liquid fraction of the aqueous foam corresponding to the volume occupied by the aqueous solution over the total volume of the foam. In the following, we first discuss how the characteristic dimension of the oil globule compared to those of the foam determines the structure of the OLF obtained. Then, we review the state of the art concerning the individual films and PBs when they are oil-laden. To do so, we organize our discussion by keeping in mind how the oil-laden foam is prepared. Indeed, as sketched in Fig. 1c), two situations can be thought of: oil can be put into contact with the foam by coming either from the gas phase enclosed in the bubbles, as observed when an oil aerosol is dispersed above an aqueous foam, or from the liquid phase if the oil is injected directly in the aqueous network of the foam, as observed when an emulsion is foamed. For these two configurations, we successively consider the different lengthscales of the foam, namely the soap films, the Plateau-Borders and the macroscopic foam scale.

3.1. Characteristic foam dimension

We first recall different features concerning the architecture of pure aqueous foams. From a fundamental point of view, foams can be seen as a dense packing of air bubbles. The structure of a foam highly depends on the liquid fraction $\varepsilon$: if $\varepsilon < \varepsilon*$, the bubbles are closely packed together and their geometry is polyhedral. In this limit of dry foam, the liquid phase is negligible and the foam is dry. The transition value $\varepsilon*$ is not well defined and is usually taken around 0.05 [1, 2]. The typical dimensions of the elements of the aqueous foam also depends on $\varepsilon$ and on $R$ the equivalent radius of the bubble, which is the typical lengthscale of the system. In the following, we introduce three different lengthscales to characterize them: $h$, the thickness of the foam films, $r_p$, the radius of passage of the PB, corresponding to the radius of the largest sphere that can flow within the PB without deforming it and $r_n$, the radius of passage of the nodes or vertexes (see Fig. 1a)). Typically, $r_p = f(\varepsilon)R$ and $r_n = g(\varepsilon)R$, where f et g dimensionless increasing function of the liquid fraction, $\varepsilon$, which are discussed into more details in the following.

Having this in mind, we now discuss the different structures of OLF that one can think of depending on the relative volume of v, the oil globule and h, $r_p$ and $r_n$ the typical dimensions of the elements of the foam. However, we stress that this classification is only approximate. Indeed, in many systems, the oil globules are not monodisperse: their distribution in volume can be wide and spans on more than one of the intervals defined in the following.

- If $v \gg r_p^2 R$, i.e. if the oil globule volume is larger than the volume of a PB, the oil globule fills more than a PB and spreads along several bubbles via the PB and the nodes. Therefore, the oil phase appears to be continuous at the scale of several bubbles. Given the length mismatch between h and $r_p$, it is unlikely that oil can be found in the adjacent foam films. We name this type of OLF as biliquid foams (Fig. 1b and Fig. 5) [51].
- If $r_p^2 R > v \gg 4/3\pi h^3$, the oil globules are dispersed in the aqueous network of the foam at the scale of a PB length. This doesn't suggest that the oil globules are spherical. Indeed, they can be deformed by the PB as can be seen in Fig. 5. Here also, we consider that the oil globules are too large to fill up the foam films. Such OLFs constitute foamed emulsions or foamulsions (see Fig. 1b) and Fig. 3) as depicted in [29, 52].

- If $v \sim 4/3\pi h^3$ or $v < 4/3\pi h^3$, the oil globules may fill up all the elements of the foam including the foam films [28].

### 3.2. Oil-laden Plateau borders

Here we discuss the geometry of an oil globule in a PB. As discussed earlier (section 3.1), an evident critical parameter is the ratio of the oil globule radius *a* to the radius of passage through the PBs $r_p$ (see Fig. 1a)[53]. Oil globules smaller than the radius of passage of PB do not deform it, while large oil globules can deform it and be deformed by the Plateau border. In the following, we recall how $r_p$ varies with the foam structure. Then, in the limit of the large oil globules, we discuss the equilibrium shape of a single large oil globule trapped inside a PB and what the spreading dynamics of a large oil volume inside a PB.

#### 3.2.1. Radius of passage of PB

The radius of passage of PB can be determined by using the Surface Evolver software [54]. For liquid fraction $0 < ? < 0.02$ and bubbles ordered in a bcc arrangement, the PB are straight slender channels and the radius of passage depends on the total liquid fraction of the foam $\varepsilon$ and R the radius of the bub- ble through $\varepsilon = f^{-1}(r_p/R) = \delta_1 (r_p/R)^2 + \delta_2 (r_p/R)^3$ with $\delta_1 = 13.8$, $\delta_2 = 145$ and *f* is the dimensionless function defined in section 3.1 [55, 53]. For a foam with a mean bubble radius R=1mm and a liquid fraction $\varepsilon$=0.01, $r_p \sim 25$ μm, while for $\varepsilon = 0.05$, $r_p \sim 50$ μm. For higher values of liquid fraction, the foam structure may change from a bcc to a fcc. Yet, an implicit function, valid for the whole range of liquid fraction, derived from the interpolation between the bcc structure at low liquid fractions ($\varepsilon < 0.1$) and the fcc structure at large liquid fractions ($\varepsilon > 0.05$) has been proposed and validated experimentally following the fate of sedimenting calibrated solid particles in an ordered foam [53].

#### 3.2.2. Small oil globules

Oil globules with $a < r_p$ aren't deformed by the PB, as shown in Fig. 3. Moreover, they appear to be densely packed. This feature has been originally explained by [12] as follows. In general, the density of oil globules is less than one. Thus, the oil globules drain slower than the aqueous phase in the PBs. This results in the accumulation of oil globule within the draining PBs. The oil globules get trapped in the PBs, because of the confinement of the PB geometry. Once trapped, the shrinking PBs squeeze the oil drops into a closely packed configuration (Fig. 3) [12, 11, 5, 8].

#### 3.2.3. Large oil globules

Neethling et al. [56] performed the numerical simulations of an oil globule elongated inside a single Plateau border. Using Surface Evolver [54], they discuss how an oil globule deforms and is, in turn, deformed by the Plateau border of the foam as can be seen in Fig. 5a). To describe these systems, the first parameter to introduce is the ratio between *a*, the equivalent spherical radius of the oil globule and $r_p$, the radius of passage of the PB. The second parameter is the ratio between the oil-water and the air water interfacial tensions: $\gamma_{ow}/\gamma_{aw}$. Indeed, when $a > r_p$, both the globule and the Plateau border become distorted. Since the shape adopted by the oil globule is a compromise between the amount of energy required to distort the Plateau border (i.e., air-water interface) and the amount of energy required to distort the globule (i.e.,oil-water interface), the shape of the oil globule changes as the interfacial tension ratio changes. Figure 5b) displays the change in shape of the oil globule as it increases in size for different interfacial tension ratios. Neethling et al. also reported that the size of the oil droplet and the interfacial tension ratio have a strong influence on the pressure exerted in the pseudoemulsion film and the related pressure difference with the Plateau border. Small droplets and high interfacial tension ratios induce a stronger pressure on the pseudoemulsion film, due to the curvature of the oil droplet. For most droplet sizes and interfacial tension ratios, the pseudoemulsion film is curved outwards (it swells the Plateau border). However when the interfacial tension ratio goes to zero, the pseudoemulsion film is oriented inwards to follow the curvature of the Plateau border (Fig. 5b)).

Such elongated shapes were also reported in experiments with a single horizontal Plateau border, obtained with a prism-shaped frame immersed in a bath of foaming solution, as can be seen in Fig 5c). The oil droplet penetrates into the PB and spreads within this liquid channel without breaking it (Fig. 2) until reaching its equilibrium length. The authors of [35] showed that in such confined channels, imbibition dynamics is governed by a balance between capillarity and viscosity as detailed in section 4.4.

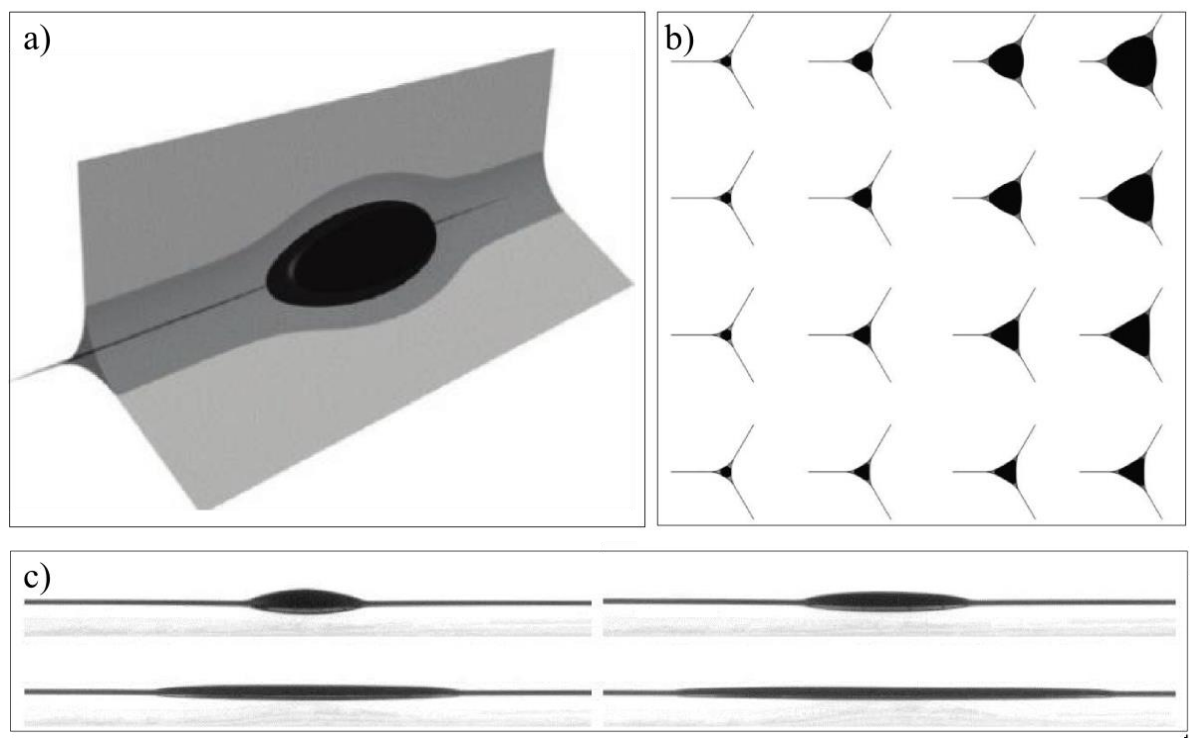

Figure 5 Oil-laden foams with $a > r_p$. a) Simulations of an oil globule in a PB. Figure reprinted with permission from [56]. Copyright, ACS, 2011 b) Cross-section views through the oil globule and the PB at their midpoints. The simulations are performed with four different interfacial tension ratios from the top to the bottom $\gamma_{ow}/\gamma_{aw}$ = 1, 0.5, 0.2 and 0.1 and from the left to right $a/r_p$ = 0.2, 0.4, 0.7 and 1.0. Figure reprinted with permission from [56]. Copyright, 2011, ACS c) Series of experimental photographs showing how an oil drop of given volume spreads along a single horizontal PB that corresponds to the horizontal line in each image. The length of the oil globule on the first image is 3 mm and the timelapse between the images is 0.2s, 3 and 10s Figure reprinted with permission from [35]. Copyright, APS, 2013.

3.3. Oil-laden Nodes

Since $r_n > r_p$, the size of the nodes usually does not impose any restriction. To the best of our knowledge, there is not any work reporting a specific role of the nodes in the OLF literature.

4. Evolution of oil-laden foams

Even if the thermodynamic conditions described in 2.1.1 are fulfilled, oil- laden foams, like their pure aqueous counterparts, are thermodynamically metastable systems, which inevitably evolve towards foam destruction and phase separation of the different components. Yet, the dynamics of evolution of OLF are even more complicated than the evolution of pure aqueous foams and exhibit non-monotonous stages where the foam destruction is extremely fast or seems to be blocked. This is all the more true for foamed emulsion for which different stages of evolution - where coarsening, coalescence and drainage are intimately intertwined - can be distinguished [14, 12]. First, a fast drainage of water from the initially formed wet foam is observed. During this period, the oil globules leave the foam films and get trapped in the neighbouring plateau borders and nodes. Due to their accumulation, (the volume fraction of oil can reach a value higher than the random close packing fraction of spheres [29]), the continuous phase can become non-Newtonian, with a finite shear modulus and eventually a yield stress [29][28]. Depending on the composition of the oil-laden foams, this can induce a reduction of the drainage rate, which consequently slows down the ageing of the oil-laden foam [12, 29], or other peculiar drainage behaviors of the foam, as described in section 4.4. However, this stage is also associated to an increase of the capillary pressure, which squeezes the globules up to a point where it exceeds the critical pressure Π (see section 2.1.1). This leads to rapid foam destruction, which occurs mainly through the disappearance of the upper layers of bubbles in which the capillary pressure is higher due to gravity effects. This complicated behavior, whose subtleness is only barely sketched in the description above, explains the wide variety of behaviors observed in the literature depending on the oil concentration [10], oil globule size [12, 57], oil density [12], surfactant concentration [29, 58], oil type [58, 18], presence of foam boosters or not [14]. As a whole, the observed behavior and the desired action of the oil globules strongly depends on the time scale of interest [14]. Thus, to get a clearer view on the specificity of the addition of oil globules on the foam evolution, we discuss separately the different mechanisms by which an OLF ages, rearranges and then collapses. We successively consider T1 topological changes, film rupture, coarsening and drainage.

4.1. T1 topological changes

Topological changes refer to the processes which modify the organization of bubbles in the foam. The T1 topological change appears when a PB shrinks to the vanishing point. This induces a metastable configuration where the neighbouring bubbles rearrange themselves in a transformation called T1 towards a new configuration of lower interfacial energy [1, 2, 3]. How oil repartition is affected by the T1 process and how the T1 dynamics is modified by the presence of oil has been investigated in [27]. In this work, the oil is injected into an elementary foam architecture, which is intentionally deformed to trigger a topological rearrangement. The authors reported how oil is redistributed within the foam architecture upon a T1, depending on the relative ratio of injected oil and water.

- For a small volume of oil globule, i.e. when the volume of oil globule is smaller than the volume of the node,, the globule is trapped and remains trapped in one node even after T1 topological changes.
- For an intermediate volume of oil globule, i.e. when the volume of oil is larger than the volume of a single node but does not spread more than one $r_p$ in the adjacent PB, oil initially trapped in one node is able to propagate to the neighbouring nodes after the T1 change. This observation shows that topological rearrangements, which naturally occur in foams when they evolve with time or when they flow, do affect the repartition of the third phase that they carry.
- • A large oil globule can spread in several nodes and PB, thus creating biliquid foams. Like for small oil volume, a T1 process does not change the oil repartition.

4.2. Film rupture

As detailed above and seen in Fig. 3, the foam films are usually not laden with oil globules except for tiny oil globules [28]. Therefore the rupture of the films of OLF is similar to their pure aqueous counterparts. More details concerning the dynamics of coalescence of pure aqueous films can be found elsewhere [59]. However, two side effects specific of the rupture of films in OLFs are worth being discussed.

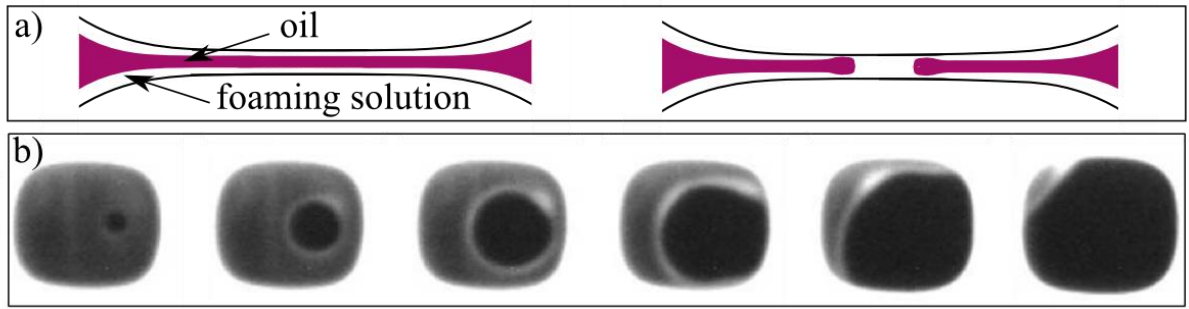

Figure 6 a) Side view of an oil laden-film formed after a T1 process. On the left, the sandwiched oil-laden film is sketched before rupture: a thin film of oil (in purple) is enclosed at the middle of a second film solely comprised of foaming solution. On the right, the oil film ruptures without breaking the aqueous film. b) Series of pictures showing from above the rupture of the oil-laden film. To enhance the contrast with water, a fluorescent dye is added to the oil. When illuminated between a blue filter and a yellow collection filter in front of the camera objective, oil appears bright on the images and water is dark. This oil-laden film, which is formed after a T1 rearrangement, breaks and retracts into the PB without breaking the aqueous film around as observed using a second camera without any light filtering (images not shown here). A hole nucleates in the oil film and grows until it reaches the Plateau borders. A rim around the hole, that becomes thicker as the radius of the hole increases, can be clearly seen. This oil film lasted typically 10 s before breaking. Interval between images: 167 ms. Reprinted with permission from [27]. Copyright, RSC, 2014.

4.2.1. Partial rupture of oil-laden fims

First, when performing their study on topological changes in OLF, the authors of [27] reported a transient unusual situation where a planar oil film is trapped within the freshly formed water film. This squeezed oil film can be stable up to a few minutes before rupturing - yet, without breaking the water film around it - as shown in Figure 6. Two important points concerning this dynamics must be highlighted. First, it nearly takes 1s to the oil film to open. This is far longer than the opening dynamics of pure aqueous films of same size, thickness and viscosity which typically open in less than 0.1s. This is due to the additional dissipation in the aqueous film enclosing the oily one. This slow dynamics modifies in turn the dynamics of the T1 process, which is an important feature for foam rheology as the scaling of local T1 dynamics should be the basis of mesoscopic models of foam flows [60, 61, 62, 63]. Note that the formation and partial coalescence of such oil laden films are not restricted to soap films held on wire frames and have also been observed at the foam scale in our laboratory [32].

4.2.2. Evolution of oil laden PB after a film rupture

The second feature of OLF films rupture is the fate of oil laden PB when an adjacent film is broken, as described in [32, 35]. To study this phenomenon, a slender oil laden PB is prepared using a wire frame as described in 3.2.3. When one foam film supporting this PB is intentionally broken, the threefold equilibrium structure of the system is modified: the direction of the two remaining film changes until they align on a single direction and merge. Concomitantly, the liquids (water and oil) are rapidly transferred from the PB to the sole soap film left (see Fig.7a)). This sudden change is followed by the fragmentation of the oil into small oil droplets embedded within the remaining soap film. This surprising feature is characteristic of the Rayleigh-Plateau instability: when the slender PB is supported by three films, the elongated oil slug is curved outward as shown in the last line of Fig. 5b). Yet, the rupture of one film induces an inversion of the topology of the system: the curvature of the oil slug turns inward, thus it breaks under the Rayleigh Plateau instability (see Fig. 7b). The experimental wavelength observed compares well with the solutions of the Rayleigh-instability problem of a viscous cylinder embedded in another viscous fluid described in [64] for oil viscosities ranging between 2.2 and 43.6 mPa.s [32]. This feature has also been observed at a larger scale in biliquid OLF, as can be seen in Fig. 7c), where several oil globules have appeared after the rupture of a film.

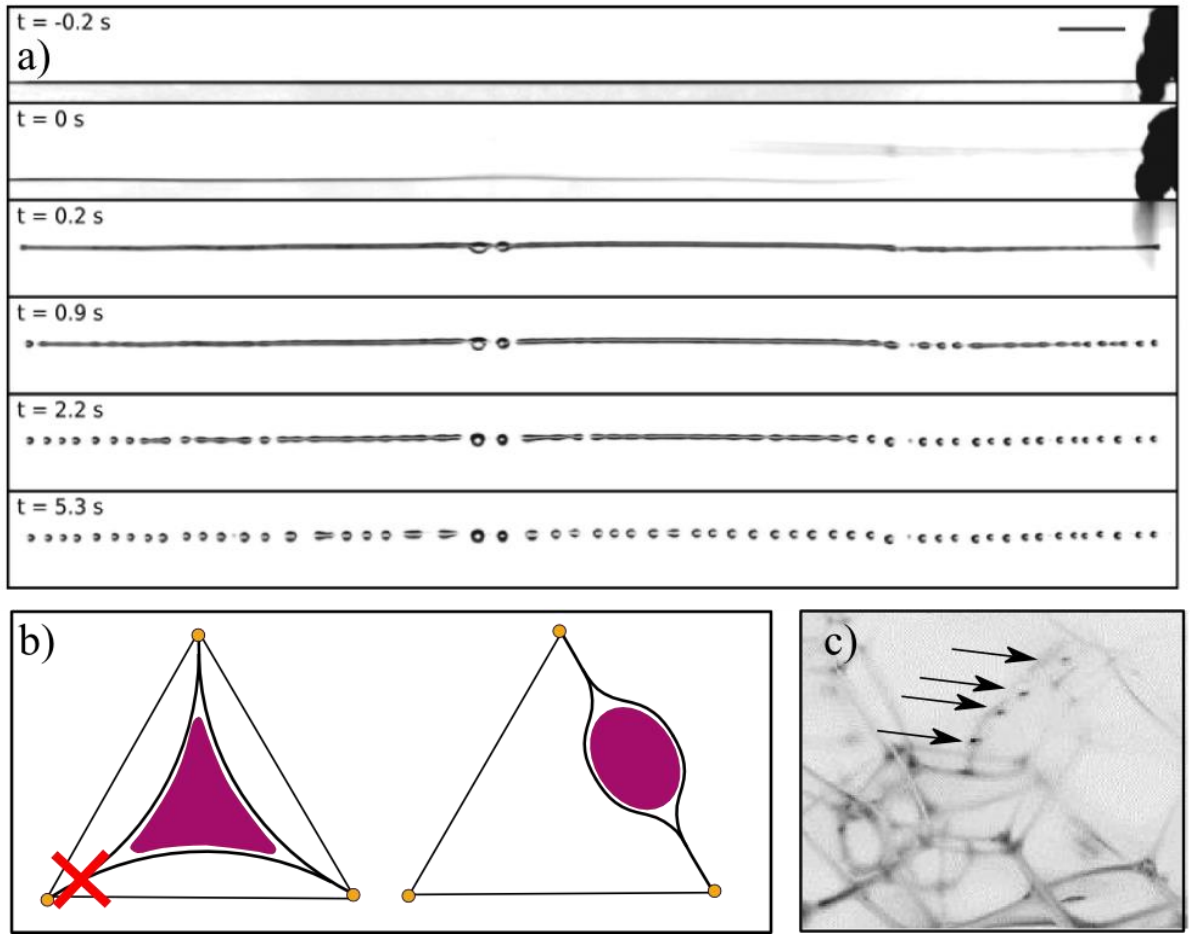

Figure 7 a) Fragmentation into droplets of an oil slug (initially comprised within a Plateau border (black horizontal line in the first image). One of the two bottom films is broken with a tissue (visible on the right side of the two first images), which leads to the transfer of the liquids from the Plateau border to the remaining film (second image). Once the oil thread is in the film, it quickly destabilizes into small globules. The bar represents 5 mm Reprinted with permission from [35]. Copyright, APS, 2013 b) Side view of a). c) Fragmentation into droplets of an elongated oil slug after the rupture of a single foam film in a biliquid OLF. The radius of the foam is typically 2 mm. To enhance the contrast with water, a fluorescent dye is added to the oil. When illuminated between a blue filter and a yellow collection filter in front of the camera objective and after inverting the contrast of the image, oil appears dark on the images and water is bright. The oil droplets are highlighted with arrows.

4.3. Coarsening of OLF

Coarsening of foams is mediated by the dissolution of gas between bubbles. This phenomenon, which has been described in great details elsewhere for pure aqueous foams [2, 3, 65, 66], is led by the existence of a surface energy between the different phases of the system. Thus, coarsening can be controlled varying either the gas solubility in the continuous phase [67, 68, 66] or the surface viscoelasticity of the air/foaming solution interfaces [66, 69, 70, 71] or the bulk viscoelasticity of the continuous phase [71]. Inspired by this constructive framework, several strategies to control the coarsening of OLF can be thought of.

In the limit where the oil globules of the foamed emulsion are smaller than the film thickness, the gas solubility could be modulated using the contrast of gas solubility between the aqueous and oil phases. Yet, as emphasized previously, even if the oil globules are initially small enough to fill up the foam films, this situation is only temporary. The foam films of free-standing foams will inevitably thin upon gravity drainage until reaching a thickness smaller than the oil globules. The latter will be expelled from the films leaving a foamed emulsion with pure aqueous films, thus explaining why this strategy - specific of foamed emulsion - has never been reported in the literature.

The control of the coarsening of OLF via a change in the surface viscoelasticity is discussed in [4]. Thermally crosslinked protein multilayers have been used to coat the bubbles of an OLF. This presumably induces an increase of the surface elasticity and viscosity and explains why air-filled emulsions based on protein capsule shells made from crosslinked cysteine-rich protein (bovine serum albumin or egg albumen) exhibit long-term stability [72]. Also, assembling larger objects at the air/foaming solution interface of the bubbles – such as close-packed layer of emulsion droplets - may be another potential mechanism for stabilization under coarsening as observed for particle laden foam [73, 74]. This has been done using rigid hydrophobin protein molecules, which are thought to behave rather like an amphiphilic Pickering nanoparticle than a classical food protein [4, 75].

Last, the viscoelasticity of the continuous phase of the foamed emulsion can also have a significant effect on the rate of change in bubble radius. This can be understood considering the scale for which the emulsion appears as a non-newtonian homogeneous fluid. In this framework, if the excess radial bulk stress tensor in the vicinity of the surface of a bubble is sufficient to counteract capillary pressure, then gas dissolution can be blocked [71, 76]. As classically observed for second order mechanical systems, viscous bulk stress tensor can only slow down the coarsening dynamics while elastic bulk stress tensor can truly blocked it. The shear modulus of concentrated emulsion typically scales as $F(\varphi)\gamma_{ow}/a$, where F is an increasing function of $\varphi$ the oil liquid fraction, $\gamma_{ow}$ is the oil/water interfacial tension and $a$ the equivalent radius of the oil droplet. Thus, to overcome the capillary pressure scaling as $\gamma_{aw}/R$, one should get at first order $F(\varphi)\gamma_{ow}/a \gg \gamma_{aw}/R$. Larger values of R/a and foamed emulsion concentrated in oil will therefore promote the formation of OLF less prone to coarsen.

4.4. Drainage

Foam drainage, which is the flow of liquid through the foam network, has been extensively studied for pure aqueous foams [61]. In terrestrial experiments, drainage can be driven by gravity and capillarity. When gravity overcomes capillarity (which is typically the case for wet foams) the foaming solution flows downward, while the gas bubble moves upwards, while in capillarity-induced drainage (where the capillary suction overcomes gravity as observed in dry foams), the foaming solution moves towards the dryer region of the foam to ensure the equilibrium of the foam. Flow in foams, either driven by capillary or gravity or both, are mainly resisted by viscous effects.

A distinctive feature of foams compared to solid porous material is that the liquid flow occurs throughout PB and nodes, which have volumes varying in space and time. Thus, the dynamics of imbibition of aqueous foaming solution into pure aqueous foam solution is ruled out by a non-linear partial differential equation called the drainage equation, which has been analyzed, compared with experimental data and solved in various situations [55, 77, 78, 79, 80, 81, 82, 83].

The addition of a third phase, oil, which is usually not perfectly density matched with the aqueous phase a priori induces more complexity to this old problem. We report below several features typical of OLF drainage.

4.4.1. Biliquid foams: dynamics of oil flow

Recent works have studied the dynamics of formation of biliquid foams following the imbibition front of oil within a dry foam using fluorescent techniques. Oil has been injected from a single point source into a single PB [35] or into a foam [45]. Other geometries where a flat oil slick is connected to a 2D or 3D aqueous foam have also been reported [30, 31]. To describe these experiments, the drainage equation, originally derived for pure aqueous foams[55, 77, 78, 79, 80, 81, 82, 83], is numerically solved taking into account gravity effect. It describes accurately imbibition experiments of both miscible (aqueous foaming solution) and immiscible (oil) liquids in dry aqueous foams provided that an effective surface tension taking into account the creation or destruction of both air/foaming solution interface and oil/foaming solution interface upon drainage is introduced [45]. As can be seen in Fig. 8, all the experimental data obtained using foams with different geometries and different oil viscosities collapse on the same master curve when the vertical front position of imbibition $z_f$ scaled by the bubble radius R is reported as a function of the time scaled by the capillary/viscous time $\tau \sim \eta_o R/\gamma$, where $\eta_o$ is the oil viscosity and $\gamma$ the effective surface tension. It can also be observed in Fig. 8 that the data for imbibition of miscible liquid is well described by the numerical solution of the drainage equation [45]. These results suggest that overall the drainage dynamics of OLF is ruled by the exact same equation than their pure aqueous counterparts. In other words, additional subtle physico-chemistrial effects such as dynamical surface tension are of second order compared to gravity, capillary and viscous effects.

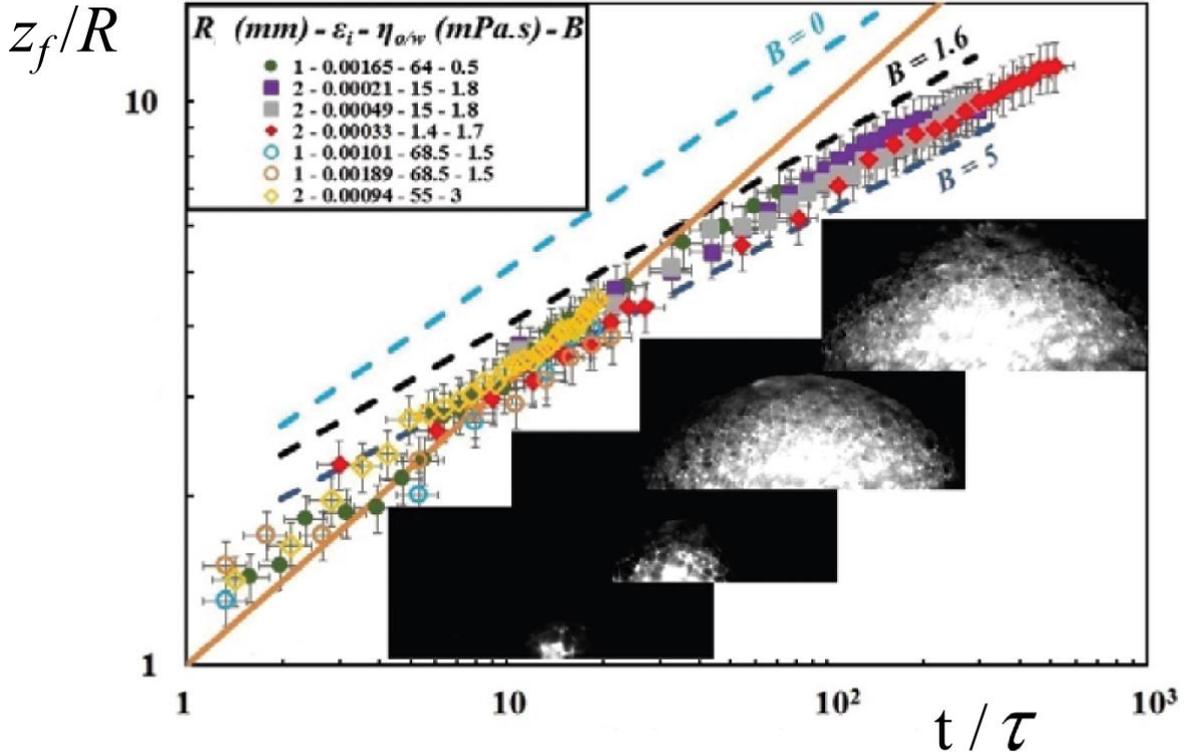

Figure 8 Vertical front position $z_f$ with respect to time in dimensionless coordinates for immiscible oils (open dots) and miscible aqueous liquids (closed dots). The experimental curves are obtained for different bubble radius R, liquid fraction of the initial aqueous foam $\varepsilon_i$, oil or water viscosity $\eta_{o/w}$ and Bond number $B = \rho g R^2/\gamma$, where $\rho$ is the density of liquid and $\gamma$ the effective surface tension taking into account the creation or destruction of air/foaming solution and oil/foaming solution during the imbibition process. Numerical solutions of the drainage equation are plotted for different values of B by the dashed lines. The self-similar power law evolution in $\tau^{1/2}$ in the no-gravity case is also shown in plain line. Reprinted with permission from [45]. Copyright, EPLA, 2016.

4.4.2. Foamed emulsion: gravity induced drainage

Foamed emulsion, where the oil globules form a dense network, usually exhibits lately very long halftime [58, 34, 29, 26, 12, 28]. This can be understood considering the non-newtonian rheology of the dense emulsion. Indeed, these dispersed materials are often viscoelasticitic and yield stress fluids. Even though, the yield stress $\tau_y$ of an emulsion or foam is not defined unambiguously [84], it scales as [62, 3, 85]:

$$\tau_y \sim \gamma_{ow}/a (\varphi - \varphi_c)^2 \qquad (4)$$

Where $\varphi$ is the oil volume fraction in the aqueous phase and $\varphi_c \sim 0.64$ the random close packing fraction. When $\tau_y \sim \Delta\rho g a$, where $\Delta\rho = \rho_w - \rho_o$, the density difference between the liquid and oil phases, the yield stress of the fluid balances gravity. Therefore, gravity induced drainage should be stopped as soon as the oil volume fraction reaches the critical value $\varphi^*$:

$$\varphi^* \sim (a^2 \Delta\rho g/\gamma_{ow})^{1/2} + \varphi_c \qquad (5)$$

For a = 200 μm, $\gamma_{ow}$ = 20 mN/m, $\Delta\rho$ = 100 kg/m 3 then $(a^2 \Delta\rho g/\gamma_{ow})^{1/2} \sim 0.04$ and the oil globules should be deformed up to a packing volume fraction slightly higher than the random close packing fraction, yet for a = 20 μm, the first term of equation 5 is negligible and $\varphi* = \varphi c$. The accumulation of small oil globules in the PB thus spontaneously leads to stable OLF due to the apparition of a yield stress in the interstitial fluid.

4.4.3. Coupling gravity drainage with shear

This stabilizing effect disappears when the OLFs are not at rest. Indeed, rapid OLF drainage has been observed under controlled shear in a direction orthogonal to gravity. The interstitial yield-stress fluid then behaves as a viscous fluid in the direction orthogonal to shear, which is that of gravity, and is thus drained. However, striking differences in the drainage dynamics have been observed with the case of static aqueous foams, namely, the independence of the velocity on the bubble size and the unexpected arrest of drainage at high liquid fraction [28]. These behaviours have been understood by considering the shear induced flow of interstitial material in the transient foam films and its coupling with the one in the foam channels [28]. Another work, coupling OLF drainage with mechanical sollicitations of OLF, by applying a force to the OLF sample, showed an acceleration of the OLF collapse [52].

5. Conclusion

We show how the physical and the mechanical properties of aqueous foams offer opportunities for the applications where one would like to find an alternative method to extract a liquid phase from confined medium, such as soil remediation and enhanced oil recovery. Foams are light, have a very little water content and offer a high specific interfacial area, which generates less waste, less energy supply and requires easier processes. We deal with the different criteria on which the stability and the collapse of foams in contact with oil depends. If stability is found with the appropriate oil-surfactant couple, one can create two types of oil-laden foams. Indeed, common examples of oil interaction with foams refer to the incorporation of oil droplets (emulsified droplets) during foam generation to generate foamed emulsion. The second type of OLF corresponds to the situation where oil appears as long slugs invading the Plateau borders at the scale of several Plateau Borders. This case can be obtained by having oil absorbed by the foam due to the capillary pressure between the Plateau border and the outer environment. However, OLF are fragile structures which evolve with time and which can lose their dual characteristics. Breaking events and drainage processes change the inner structure of the OLF.

A lot of works have been carried out in the field of OLF. Yet, many relevant issues are still unexplained. For instance, as for purely aqueous foams, we are far from being able to make the link between the macroscopic scale of thermodynamics, for which the established theoretical framework is in good agreement with the experimental observations, and the molecular scale characteristic of the structure and composition of the surfactant molecules. The destabilization of a Plateau border filled with oil starts by the breaking of the pseudoemulsion film which is the thin aqueous film between the air phase and the oil phase. This can only happen if the entry of the oil phase at the air-water interface overcomes an entry barrier related to the pressure required to thwart the disjoining pressure in the pseudoemulsion film. The relationship between the entry barrier and the well-recognized static thermodynamical coefficients is still an open question. Another field where the physics of OLF remains partially unveiled concerns the coupling between the different ageing mechanisms, which is nearly unexplored.

Acknowledgements

We acknowledge F. Rouyer, O. Pitois, A.L. Biance, I. Cantat, A. Delbos, Y. Peysson, S. Youssef, B. Laborie, E. Rio and K. Piroird for fruitful discussions. This work has benefited from two French government grants managed by ANR within the frames of the national program Investments for the Future (ANR-11-LABX-022-01) and of the young researcher program (ANR-11-JS09-012-WOLF)